\documentstyle[twocolumn,prl,aps,psfig]{revtex}

\begin{document}

\tightenlines

\twocolumn[\hsize\textwidth\columnwidth\hsize\csname @twocolumnfalse\endcsname

\title{Resistivity of Mixed-Phase Manganites}

\author{Matthias Mayr$^1$, Adriana Moreo$^1$, Jose A. Verg\'es$^2$,
Jeanette Arispe$^1$, Adrian Feiguin$^1$,
and Elbio Dagotto$^1$}

\address{$^1$ National High Magnetic Field Lab and Department of Physics,
Florida State University, Tallahassee, FL 32306, USA}

\address{$^2$ Instituto de Ciencia de Materiales de Madrid, 
Cantoblanco, E-28049, Madrid, Spain.}

\date{\today}
\maketitle

\begin{abstract}

The resistivity $\rho_{dc}$ of manganites  is studied using a 
random-resistor-network, based on phase-separation between metallic and
insulating domains. When percolation occurs, both as 
chemical composition and temperature vary, results in good
agreement with experiments are obtained. Similar conclusions are reached
using quantum calculations and microscopic
considerations. Above the Curie temperature, it is argued that
ferromagnetic clusters should exist in Mn-oxides.
Small magnetic fields induce large $\rho_{dc}$ changes 
and a bad-metal state with (disconnected) insulating domains. 
\end{abstract}

\vskip1pc]
\narrowtext

The study of manganites is one of the main
areas of research in Strongly Correlated Electrons\cite{tokura}.
Three main reasons have triggered this wide interest:
(1) The low-bandwidth
materials have unexplained transport properties. They are
insulators
at room temperature, changing into bad metals
at low temperatures. A sharp peak in 
the resistivity $\rho_{dc}$ appears at the ferromagnetic (FM)
transition. Small magnetic fields turn the
insulator into a metal, 
with a ``Colossal'' Magneto-Resistance (CMR).
(2) The phase diagram T-x (T=temperature, x=hole density) 
is rich, with complex spin, charge, and orbital order. 
(3) Mn-oxides have intrinsic 
inhomogeneities in most of the T-x plane
even in single crystals\cite{cheong,moreo}. 

This challenging behavior has been addressed by previous theoretical
studies. Regarding item (2), the various 
FM, antiferromagnetic (AF), 
orbital-ordered, and charge-ordered (CO) phases 
have already emerged from simulations and mean-field
approximations\cite{yunoki1}.
Regarding item (3), phase separation (PS) has been proposed
to explain the inhomogeneities. PS can be:
(a) electronic, with nanometer-size 
clusters\cite{moreo}, or
(b) structural, where disorder can induce up to micrometer-size
clusters and percolation,
when influencing on first-order transitions\cite{random}.
However, the explanation of transport,
item (1), is more complicated since estimations
of $\rho_{dc}$ are notoriously 
difficult. In addition, 
the prominent inhomogeneities 
of Mn-oxides\cite{cheong} have not been incorporated into realistic 
$\rho_{dc}$ calculations\cite{comm10}.
The behavior of $\rho_{dc}$ remains unexplained, 
although it is central to manganite physics.

Our goal in this paper is to present a rationalization of the 
$\rho_{dc}$ vs T curves of Mn-oxides based on the currently prevailing
phase-separated/percolative framework for these compounds. In this
context, items (1) and (3) above are closely related.
The analysis necessarily involves 
phenomenological considerations, since percolation
cannot be addressed
on sufficiently large lattices using
accurate microscopic models. However,
the $\mu$-meter clusters in  experiments\cite{cheong} strongly suggest
that a coarse-grain approach
should be sufficient. In addition, results of microscopic
calculations presented below are
consistent with those of the macroscopic approach.

The main concept introduced here is summarized in Figs.1a-b.
The manganite state in the CMR regime 
is assumed to be percolative, with 
metallic filaments across the sample (Fig.1a). 
Percolation indeed occurs in
models\cite{random}, and in 
many experiments\cite{cheong}. The insulating and metallic
(percolative) regions are assumed to have 
resistances $\rm R_I$(T) and $\rm R^{per}_M$(T),
respectively, as sketched in Fig.1b.
$\rm R^{per}_M$ is large at T=0 due to the complex shape of
the conducting paths, and grows with T as in any metal, eventually
diverging when the percolative path melts with increasing T. Note that
at room temperature $\rm R_I$$<$$\rm R^{per}_M$
and, thus, most of the conduction in this regime
occurs {\it through the insulator}. 
On the other hand, $\rm R_I$ is so large at low-T 
that current can only flow through the percolative paths. This
suggests a simple two parallel
resistances description (Fig.1b), where a
peak in the effective resistance at intermediate T is natural.

To substantiate this idea, first consider
results obtained using a {\it random-resistor-network}
that mimics the prominent FM-CO mixtures\cite{cheong}
in Mn-oxides in the CMR regime.
Two-dimensional (2D) and three-dimensional (3D) 
square and cubic clusters are used, with link resistivities
randomly selected as
metallic ($\rho_M$) or insulating ($\rho_I$), 
with a fixed metallic fraction $p$ 
(which in, e.g., $\rm (La_{5/8-y} Pr_y) Ca_{3/8} 
Mn O_3$ (LPCMO) it is proportional to the 
amount of La \cite{cheong}). The lattice spacing of this effective network
is comparable to the FM or CO domain size, 
much larger than the Mn-Mn distance.
The actual values of $\rho_M$ and $\rho_I$
vs T were directly taken from LPCMO data\cite{cheong}
(y=0.00 and 0.42, respectively), and for simplicity they are used
in both 3D and 2D clusters. Other materials were tried
and the analysis below does not depend qualitatively 
on the reference compounds. 

The Kirchoff equations for the
network were solved iteratively on large clusters using well-known 
techniques\cite{kirk}, and the net resistivity was found. 
Typical results are shown in Fig.1c. 
Only the limiting cases $p$=0 (all insulator) and 1 (all metal) are
taken from experiments.
As expected, a percolative regime exists 
between $p$=0.4 and 0.5, 
where $\rho_{dc}$(T=0) 
is as large as in LPCMO and other materials\cite{comm1}. 
$\rho_{dc}$ has insulating behavior at room-T,
even for $p$ as high as 0.65, while at low-T a (bad) metallic
 behavior is observed.
A broad peak appears at intermediate
T's and $p$'s. Similar results exist in 3D (inset of
Fig.1c). 

\begin{figure}
\psfig{figure=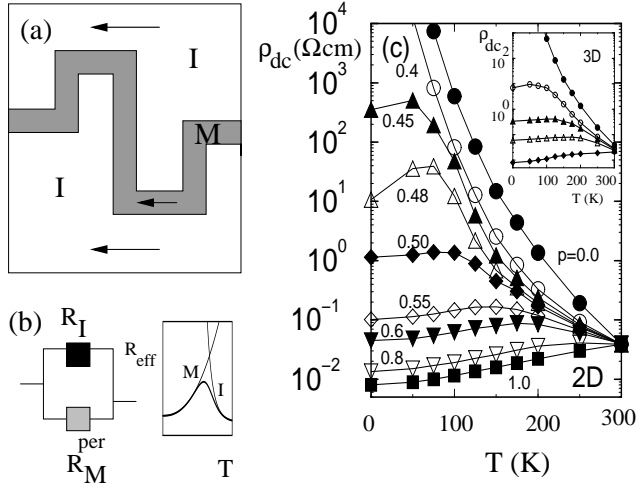,height=6.5cm,angle=0}
\vspace{0.4cm}
\caption{ (a) Mixed-phase state near percolation. Arrows indicate
conduction through insulating or metallic regions depending
on T (see text). 
(b) Two-resistances model for Mn-oxides. 
Effective resistance $\rm R_{eff}$ vs T (schematic) 
arising from the parallel connection
of metallic (percolative) $\rm R^{per}_M$ and insulating 
$\rm R_I$ resistances. 
(c) Net resistivity $\rho_{dc}$ of a 
100$\times$100 cluster vs T, 
at the indicated
 metallic fractions $p$. Inset: Results for a 20$^3$ cluster 
with (from top) $p$=0.0, 0.25, 0.30, 0.40 and 0.50. 
In both cases, averages over 40 resistance configurations
were made (errors the size of the points). The $p$=1 and 0 limits are from
Ref.\protect{\cite{cheong}}. 
Results on 200$\times$200 clusters (not shown) indicate that
size effects are negligible. }
\label{Figure_1}
\end{figure}

It is remarkable that Fig.1c
is already in good qualitative agreement with some Mn-oxide
experiments, such as for
$\rm La_{0.96-y} Nd_y K_{0.04} Mn O_3$ \cite{mathieu},
or even non-manganite materials, 
such as $\rm Ca Fe_{1-x} Co_x O_3$ where an AF-FM 
competition occurs\cite{kawasaki}. However, many manganites present 
a more pronounced $\rho_{dc}$ peak 
at intermediate T's. To reproduce this feature,
it is necessary to    
introduce a percolative process not only as $p$ (or x) varies, but
also as T changes. This is reasonable since the
metallic component triggered by ferromagnetism is sensitive
to T, and the FM clusters 
shrink in size as T increases.
This proposal was tested qualitatively using two models:
(i) the Random Field Ising model (RFIM), that
describes the disorder-induced PS\cite{random}, 
with spin up and down crudely representing the competing 
metal and insulator,
and (ii) the well-known one-orbital model\cite{moreo} (with
parameters t, $\rm J_H$, and $\rm J'$ representing the 
$\rm e_g$-hopping, Hund coupling, and Heisenberg exchange 
among $\rm t_{2g}$-spins, respectively). The latter is
supplemented by a term $\sum_i$$\phi_i$n$_i$,
with $\phi_i$ randomly
taken from [-W,W] and n$_i$ the on-site density at
site $i$. This disorder 
generates coexisting clusters near first-order FM-AF transitions,
as explained in Ref.\cite{random}. 

In Fig.2a, it is shown
a portion of a typical Monte Carlo (MC) simulation 
of the RFIM at T=1.6 
on a 500$\times$500 cluster, with fixed random fields taken from 
[-1.0,1.0] 
(J=1 is the FM Ising coupling). These parameters are the
same as in Ref.\cite{random}. Three main
domains were found: spins-up (black), 
spins-down (white), and 
regions with a small spin expectation value (grey).
The generation with T
of these paramagnetic (PM) areas in the surface of the up and down
domains weakens the percolative tendencies of the RFIM: for
example, in Fig.2a domains 
connected at T=0, become disconnected at finite-T.
Similar behavior occurs in the microscopic
one-orbital case, as shown in Fig.2b 
for a disordered configuration with
percolative features\cite{comm11}.
Increasing T from $\sim$0.0 to 0.05t
decouples the two FM regions. FM, AF, and PM regimes 
dominate at finite T, as in the RFIM.
The T=0.05t Drude weight (not shown) is much
smaller than at T$\sim$0.0.

\begin{figure}
\psfig{figure=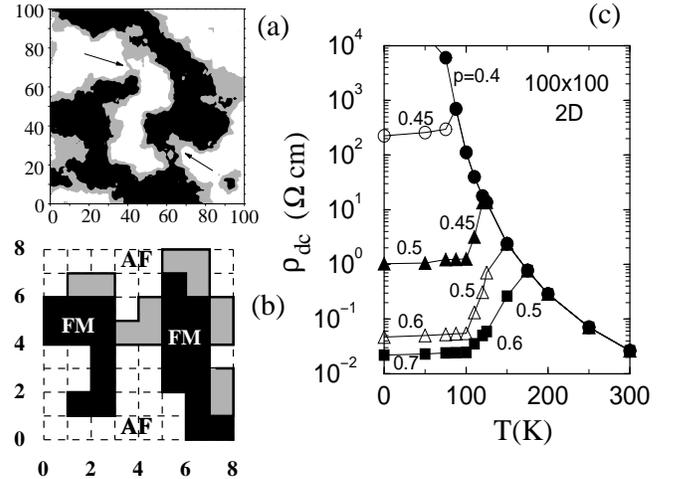,height=6.5cm,angle=0}
\vspace{0.4cm}
\caption{ (a) MC results for the Random Field Ising Model at
T=1.6 (J=1), and a fixed set of random fields taken from [-1.0,1.0]. 
Shown is a 100$\times$100 subset of a 500$\times$500 cluster
with periodic boundary conditions (PBC). 
Black, white, and grey denote regions with Ising
spin mean-value $\langle$$\rm \sigma$$\rangle$ in the intervals
(0.66,1.0), (-1.0,-0.66), and (-0.66,0.66), respectively. The arrows indicate
regions where percolation occurs as T decreases.
(b) MC results for the one-orbital model on a PBC 8$\times$8 cluster,
$\rm J_H$=8, $\rm J'$=0.0, t=1, density $\langle$n$\rangle$=0.875,
and a (fixed) realization of disorder (W=0.5) with 
percolative characteristics at T$\sim$0. Black and  white
denote FM and AF regions, respectively. 
Grey domains are FM at T$\sim$0, but the nearest-neighbor (NN) spin
correlations become much smaller (PM) at T=0.05.
(c) Net $\rho_{dc}$ of a 100$\times$100 cluster as in Fig.1c,
but with $p$ changing with T (representative values indicated). 
Averages over 40 resistance configurations
are shown.}
\label{Figure_2}
\end{figure}

To incorporate the indications of
T-induced percolation (Figs.2a-b) in
the phenomenological approach, 
a T-dependent metallic fraction $p({\rm T})$ is needed. $p$ must
decrease as T grows, should vary rapidly near the Curie temperature 
$\rm T_C$ as the magnetization does,
but otherwise its T-dependence is unknown. Fortunately, 
the qualitative results using several functions are similar,
and a typical case is shown in Fig.2c.
The $\rho_{dc}$'s obtained by this simple procedure now
clearly resemble those found in experiments, 
with a robust peak at intermediate
T's, moving toward lower T's as the system becomes
more insulating. This agreement with experiments is
unlikely to be accidental, and justifies {\it a posteriori} 
our assumptions. 
Note that if our approach is correct, consistency requires
that in Mn-oxides above $\rm T_C$ there should
exist (disconnected) FM clusters on an insulating matrix,
since $p$ does not drop abruptly to 0 at $\rm T_C$. 
A new temperature scale T$^*$ is predicted,
with those FM clusters existing in the range 
$\rm T_C$$\leq$T$\leq$T$^*$. The density of states in this regime
likely has a $pseudogap$, based on previous investigations of
mixed-phase states\cite{moreo,random}. 

Consider now nanometer-scale
clusters. Here quantum effects cannot be neglected.
However, the problem is
still too difficult to be treated microscopically, and
an effective description is needed. For this purpose, 
instead of a resistor network,
a 3D lattice model with NN
electron hopping
(and zero chemical potential) is here used,
with link hopping amplitudes randomly selected to be either
``metallic'' ($\rm t_M$) or ``insulating'' ($\rm t_I$), representing
effective hoppings through the nanoclusters
with 1 (0) corresponding to the FM (AF) regions of the
microscopic model at large $\rm J_H$. Such ``lattice of quantum wires''
has been used before to study quantum percolation\cite{luck}.
The cluster conductance $C$ (in $\rm e^2/h$ units)
is calculated using the Kubo formula within a Landauer 
setup\cite{verges}, connecting the finite clusters
to semi-infinite leads using an infinitesimal
voltage drop. The self-energy of these leads is known
exactly\cite{verges}, and the lead on the, e.g., right
is used for an iterative calculation of the self-energy from right to
left through the
cluster, supplemented by a self-energy matching at the left end.
With the Green function obtained by this procedure, 
$C$ is evaluated using known
formulas\cite{verges}.
Since this is $quantum$ percolation, the study 
is only in 3D (2D localization leads to zero
conductivity).

The hoppings $\rm t_M$ and  $\rm t_I$ are not available from
experiments. However, $\rm t_M$ should
decrease with increasing T following the FM-phase
magnetization,
while $\rm t_I$ increases
with T (since, e.g., the zero-conductivity T=0 AF
configuration disorders as T grows).
$C$ was obtained on up to 20$^3$ clusters using the
$\rm t_M$(T) and $\rm t_I$(T) in the inset of Fig.3a, but the 
results do not depend qualitatively on the particular functions
used, as long as $\rm t_I$ changes rapidly with T near room-T,
as $\rho_I$ does in experiments.
The metallic fraction
$p$ in Fig.3a was made T-dependent as in Fig.2c, and the critical
percolation at T=0 is expected to be located near
$p_c$$\sim$0.45 \cite{luck}. Results
are shown in Fig.3a\cite{comm3}. Once again 
a remarkable qualitative agreement with experiments is obtained,
suggesting that both nano- and micro-meter clusters lead to similar
results. 

\begin{figure}
\psfig{figure=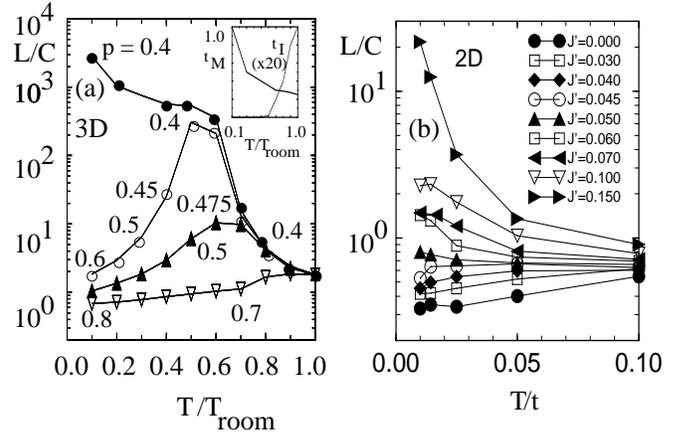,height=6.0cm,angle=0}
\vspace{0.4cm}
\caption{ (a) Inverse conductance of a tight-binding 
(NN sites) 20$^3$ (L=20) cluster with hoppings
$\rm t_M$ and $\rm t_I$. Temperature is considered through changes
in those hoppings, shown in the inset. $p$ is the fraction of
metallic links, and representative values are indicated. The chemical
potential was set to 0.
(b) Inverse conductance of the half-doped
one-orbital model on a PBC 8$\times$8 (L=8) cluster,
and $\rm J_H$=$\infty$ (t=1), varying $\rm J'$. Disorder strength is
$\Delta$=0.1. Averages over 15 disorder configurations are shown.
Error bars in Figs.3b and 4a are of the order of the oscillations of
each curve.}
\label{Figure_3}
\end{figure}

For completeness, $C$ was also 
calculated using microscopic models on small lattices. 
A MC simulation of the one-orbital model on an
 8$\times$8 PBC cluster  
and density x=0.5 
was performed. From previous work\cite{random}, it is
known that a metal-insulator
first-order transition
occurs at $\rm J'_c$$\sim$0.07 (if $\rm J_H$=$\infty$). 
Disorder is introduced such that in the
NN-sites link $\langle$ij$\rangle$ the hopping is
$\rm t_{ij}$=1+$\delta_{ij}$ and Heisenberg coupling is
$\rm J'_{ij}$=$\rm J'$(1+$\delta_{ij}$), where $\delta_{ij}$
is randomly taken from [-$\Delta$,$\Delta$],
and $\rm J'$ is uniform. This disorder
makes the transition
continuous\cite{random}. The MC procedure generates $\rm t_{2g}$-spin
configurations from which NN-sites
effective hoppings can
be calculated (as in double exchange models). 
These hoppings are used to evaluate $C$ (Fig.3b). Due to the disorder, 
$C$ interpolates smoothly from metal 
to insulator varying $\rm J'$ (otherwise a discontinuous transition occurs). 
$C^{-1}$(T$\sim$0) can be very large, but finite, if
the appropriate value of $\rm J'$ is selected\cite{comm30}.
In this respect the result has clear similarities 
with those of the
macroscopic approach. 
However, 
the full shape of the experimental $\rho_{dc}$ curves
is difficult to reproduce with microscopic models on small clusters
where percolation cannot be studied.
Nevertheless, for the (few) disorder configurations with percolative-like
characteristics found on small systems
(as in Fig.2b), the associated $C$ 
vs T has a broad maximum at a finite T.

Consider now nonzero magnetic fields ($h$). In the
random-network model mimicking coexisting FM-AF regions,
a small $h$ will increase the 
FM fraction $p$ by a concomitant small amount. 
However, near percolation tiny modifications in $p$ 
can induce large $\rho_{dc}$
changes, as shown in Fig.2b where a 5\% modification in $p$ at low-T
can alter $\rho_{dc}$ by two orders of magnitude. In the 
percolative regime,
``small'' perturbations can drastically change the conductivity.
This analysis predicts that the
metallic state reached from the insulator with
magnetic fields is $not$ homogeneous but must still have
a substantial fraction of insulating clusters. This is consistent
with the experimental large $\rho_{dc}$(T=0) of such a state.

\begin{figure}
\psfig{figure=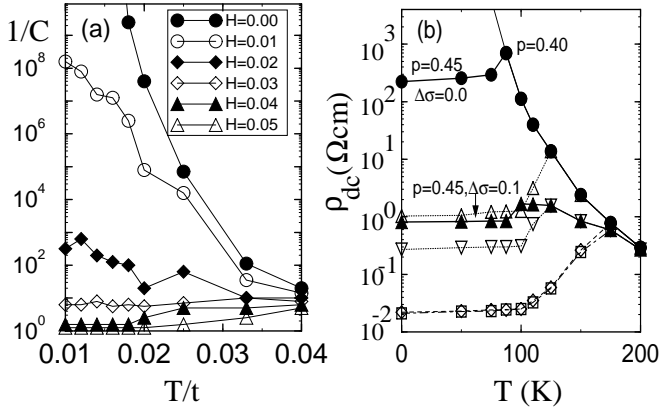,height=5.9cm,angle=0}
\vspace{0.4cm}
\caption{ (a) Inverse conductivity of the half-doped one-orbital model 
on a 64-site chain, with $\rm J_H$=$\infty$, $\rm J'$=0.14, t=1, 
and $\Delta$=0.03, 
varying a magnetic field as indicated. The data shown corresponds to
one disorder configuration, but results with other configurations are similar.
(b) Effective resistivity of a 100$\times$100 network of
resistances. Results at $\Delta \sigma$=0.0 (full circles, open triangles,
and open squares starting
at T=0 with $p$=0.45, 0.50 and 0.70, respectively) are the same as found
in Fig.2c. Full triangles, inverse open triangles, and diamonds, correspond
to the same metallic fractions, but with a small addition to the insulating
conductivity ($\Delta \sigma$=0.1 ($\Omega$cm$)^{-1}$), to simulate the
effect of magnetic fields (see text).}
\label{Figure_4}
\end{figure} 

Another effect contributes to this phenomenon. 
It exists even on chains
where percolation does not occur, and it is illustrated in 
Fig.4a where $C^{-1}$ is shown using the microscopic
half-doped
one-orbital model  at $\rm J'$=0.14 where 
the system is at a FM(metal)-AF(insulator) 
transition, the latter with the periodic spin structure 
$\uparrow \uparrow \downarrow \downarrow$,
as shown in Ref.\cite{random}. The field is introduced as
$h$$\rm  \sum_i M^z_i$, where $\rm M^z_i$=$\rm s^z_i + (3/2)S^z_i$,
with $\rm S^z_i$ the $\rm z$-component of the classical $\rm t_{2g}$-spin
at site $\rm i$ with norm 1,
and $\rm s^z_i$ the spin of the mobile electron at the
same site.
Disorder in the hopping and $\rm J'$ (as in Fig.3b) of strength
$\Delta$=0.03
produces coexisting FM-AF clusters\cite{random}.
The nearly perfect AF-links at low-T induce a huge $C^{-1}$ at
$h$=0. However,
field modifications of 
just 0.01t ($\sim$17 Tesla, if $\rm t$=0.2eV)
produce dramatic changes in $C^{-1}$ at low-T (Fig.4a). 
The resulting $C^{-1}$ at $h$$\neq$0.0 is still large,
but compared with $C^{-1}$($h$=0) the effect is notorious.
An analysis of the spin correlations vs $h$ shows that
these large resistance changes mainly originate in the 
$AF$ regions,
since small fields produce a small spin canting and concomitant
small conductivity, creating a $valve$ effect between metallic
domains. 
In real manganites, a relatively modest
$\rho_{dc}$ change in the insulating regions could
contribute appreciably to the large MR.
To simulate this effect, $\rho_I$ of the 2D network of Fig.2c was
slightly modified ($\Delta$$\sigma_{dc}$=$0.1$$(\Omega cm)^{-1}$), with
 $\rho_M$ untouched.
The resulting $\rho_{dc}$ changes 
(Fig.4b) are indeed large at low-T, comparable to those 
obtained changing $p$ by a few percent.

Summarizing, the manganite $\rho_{dc}$ was studied within the
PS-framework using a semi-phenomenological
approach. At room-T,
conduction predominantly through the insulating regions leads
to d$\rho_{dc}$/dT$<$0,
while at low-T the metallic filaments carry the
current. The large MR induced by magnetic fields
is caused by small changes in the
metallic fraction $p$ and/or in the
conductivity of the insulator, effects
which severely affect transport near percolation. 
Our approach provides a simple explanation of the CMR effect,
without invoking polaronic or Anderson localization concepts, and
independently of the origin (Coulomb vs Jahn-Teller) 
of the competing phases.

The authors thank S. Yunoki for the programs used for
Figs.2b-3b-4a, and
NSF (DMR-9814350), Comision Interministerial de Ciencia y Tecnologia
de Espa\~na (PB96-0085), MARTECH, CDCH (Univ. Central de Venezuela),
and Fundaci\'on Antorchas for support.
\vspace{-0.6cm}
%------------references--------------------------------

%-------------------------figure captions-----------------------


\begin{references}
\vspace{-1.6cm}

\bibitem{tokura} 
Y. Tokura and N. Nagaosa, Science {\bf 288}, 462 (2000).

\bibitem{cheong} M. Uehara {\it et al.},
Nature {\bf 399}, 560 (1999). See also
M. Ibarra, and J. De Teresa, JMMM {\bf 177-181}, 846 (1998); 
M. F\"ath {\it et al.}, Science {\bf 285}, 1540 (1999); 
D. Louca and T. Egami, Phys. Rev. B{\bf 59}, 6193 (1999).


\bibitem{moreo} 
A. Moreo {\it et al.}, Science {\bf 283}, 2034 (1999).
See also S. Yunoki {\it et al.}, Phys. Rev. Lett. {\bf 80}, 845 (1998).

\bibitem{yunoki1} S. Yunoki {\it et al.}, 
Phys. Rev. Lett. {\bf 81}, 5612 (1998);
{\bf 84}, 3714 (2000);
T. Hotta {\it et al.}, {\it ibid} {\bf 84}, 2477 (2000);
J. van den Brink {\it et al.}, {\it ibid} {\bf 83}, 5118 (1999).



\bibitem{random} A. Moreo {\it et al.}, Phys. Rev. Lett. {\bf 84}, 5568 (2000).


\bibitem{comm10} 
Localization to understand  the 
insulator above $\rm T_C$ was
studied in L. Sheng {\it et al.}, Phys. Rev. Lett.
{\bf 79}, 1710 (1997). However, V. Smolyaninova {\it et al.},
cond-mat/9903238,
showed that the metal-insulator transition of $\rm La_{0.66}Ca_{0.33}MnO_3$
is not due to Anderson localization.


\bibitem{kirk} S. Kirkpatrick, Rev. Mod. Phys. {\bf 45}, 574 (1973).

\bibitem{comm1} Since $\rho_I$(T$\sim$0) of 
y=0.42 LPCMO is finite, the net $\rho_{dc}$(T=0) does not diverge
below the naively expected 
critical point $p_c$=0.50 of a simple 2D network with $\rho_I$=$\infty$.

\bibitem{mathieu} R. Mathieu {\it et al.}, cond-mat/0007154.

\bibitem{kawasaki} S. Kawasaki {\it et al.}, J. Phys. Soc. Jpn. {\bf
67}, 1529 (1998).

\bibitem{comm11} 
``Percolative'' configurations are rare in small clusters.

\bibitem{luck} Y. Avishai and J. M. Luck, Phys. Rev. B{\bf 45}, 1074 (1992).

\bibitem{verges} J. A. Verg\'es, cond-mat/9905235. See also S. Datta,
{\it Electronic Transport in Mesoscopic Systems}, Cambridge University
Press, Cambridge, 1995, and M. Calderon, J. Verg\'es, and L. Brey,
Phys. Rev. B{\bf 59}, 4170 (1999). 



\bibitem{comm3} It was verified
by a finite-size-scaling analysis that the conductances shown in
Fig.3a are in the ohmic regime.


\bibitem{comm30} Similar results
were obtained using the two-orbital model\cite{yunoki1} in one-dimension,
near a FM-AF transition\cite{random}.




\end{references}
\end{document}